\documentclass[manuscript]{aastex}
\usepackage{natbib}

\usepackage{amsmath}

\slugcomment{Not to appear in Nonlearned J., 45.}

\shorttitle{Optical Cavities}
\shortauthors{De la Luz et al.}

\begin{document}
\title{The Chromospheric Solar Millimeter-wave Cavity, as a Result of the Temperature Minimum Region}

\author{Victor De la Luz\altaffilmark{1,2}, Jean-Pierre
  Raulin\altaffilmark{3},  Alejandro Lara\altaffilmark{2}}
\altaffiltext{1}{Instituto Nacional de Astrofisica, Optica y Electronica,
  Tonantzintla, Puebla, Mexico, Apdo. Postal 51 y 216, 72000.}
\altaffiltext{2}{Instituto de Geof\'isica, Universidad Nacional Aut\'onoma de
  M\'exico, M\'exico, 04510.}
\altaffiltext{3}{CRAAM, Universidade Presbiteriana Mackenzie, S\~ao Paulo, SP, Brasil, 01302-907.}

\begin{abstract}
We present a detailed theoretical analysis of the local radio emission 
at the lower part of the solar atmosphere.
To accomplish this, 
we have used a numerical code to simulate the emission and transport of
high frequency electromagnetic waves 
from $2$ GHz up to
$10$ THz. As initial conditions we used 
VALC, SEL05 and C7 solar chromospheric models.
%
In this way, the generated synthetic spectra allows us 
to study 
the local emission and absorption processes with  high resolution in both 
altitude and frequency.

Associated with the temperature minimum predicted by these models we found
that the local optical depth at millimeter wavelengths remains
constant, producing an optically thin layer which is
surrounded by two layers of high local emission. We
call this structure the Chromospheric Solar Millimeter-wave Cavity (CSMC). 
The temperature profile which features the Temperature Minimum layers and the
following Temperature rise produces the CSMC phenomenon. 

The CSMC show the complexity of the relation between the theoretical 
temperature profile and the observed brightness temperature 
and may help to understand the dispersion of the observed brightness 
temperature in the millimeter wavelength range.

\end{abstract}


\section{Introduction}

The necessity of a two component (hot and cold) model to explain the
chromospheric 
solar
observations  was established by \cite{1949MNRAS.109..298G}.
A low temperature 
zone
in the chromosphere was inferred by observations  
in different regions of the solar spectrum 
in both,  line-emission (e. g.:  
{H and K of {\it Ca\textsc{ii}} in the Visible; 
 {\it Mg} {\it h} and {\it k} resonance lines in the   UV) and  continuum  (e.g.; the
 135-168 nm in 
UV  and the 33-500 $\mu$m in microwave ranges). 
Moreover, by observing emission lines  
({\it Fe}, {\it Si}, {\it K1} of {\it Ca\textsc{ii}}, and {\it k1} of {\it  Mg})  at 
the disk center,  
the height of the low temperature region was set  
at the lower part of the  chromosphere, 
 very close 
to the photosphere \citep{1970ApJ...161..713A}.

A considerable advance in the understanding of the chromosphere was
obtained by the so called VAL models. Using these models and 
 based on 
adequate spatial
resolution observations,    
\cite{1973ApJ...184..605V,1976ApJS...30....1V,1981ApJS...45..635V}  
found   
 that a minimum of temperature can be modeled  by  
assuming a 
plane-parallel atmosphere in hydrostatic equilibrium and  considering   
Non-Local Thermodynamic Equilibrium (NLTE). 
After that, 
\cite{1990ApJ...355..700F} improved the VAL models by including
several updates (both observationally and theoretically); the most important
being 
the ambipolar diffusion 
which
removed the ``plateau'' of temperature in the high chromosphere.

The most recent and advanced model is the so called ``C7 model'' 
\citep{2008ApJS..175..229A} which  also follows the common assumptions
of \cite{1973ApJ...184..605V}. 
In particular, this model 
predicts that the continuum  emission at $500$ $\mu$m  
comes from  
the region where the temperature reaches its minimum value. 
The temperature minimum
region  
has been set by the mentioned   
semi-empirical models 
at heights ranging 
between 400 and 600 km above the photosphere \citep{1978SvA....22..345K,
  1981SoPh...69..273A,1981ApJS...45..635V,1990ApJ...355..700F,2008ApJS..175..229A}.
In the literature we also found purely empirical  models, for example the SEL05 model
\citep{2005A&A...433..365S} that 
closely matches
the observations made by \cite{1993ApJ...403..426E,1976SoPh...48...41K,1973SoPh...31..319B,1973SoPh...28..409L}.


Despite the fact that 
the chromosphere (where reaches the temperature minimum)
plays a fundamental role in the dynamics 
and heating of the upper atmosphere (acting as the interface between the
lower photospheric layers where $\beta$, the ratio between the dynamic and magnetic pressures is
much higher than one and the 
higher
dynamic
atmosphere where $\beta << 1$),
this layer  
is not well understood. 
Recently a renewed interest on  this fundamental layer has grown due to
the advent of better observational  and reconstruction technics applied
to ground based  as well as space borne telescopes 
giving  access to
high 
space and time
resolution 
observations of the solar chromosphere.

In this context, the analysis of the aforementioned models predicts that the 
millimeter and 
sub-millimeter (7.5 to 0.75 mm) quiescent solar
emission comes from  
a layer in the 
lower
chromosphere
related to the 
temperature minimum 
\citep{1981ApJS...45..635V}.  

It is well established that 
the height and  width of 
a radio emitting 
layer  
can be characterized using 
radiative transport models 
\citep{2008ApJS..175..229A}. However, 
in general
these analysis 
consider 
only 
the upper layer (or final boundary) of the emission region,    
i.e. the layer where the atmosphere becomes optically thick.    
Therefore,  the  
knowledge and understanding 
of the  
local emission produced at  lower layers
remains
neglected.
Note that the fact that these regions are not easily  accessible to radio
telescopes, does not prevent us to study the possible physical processes
occurring inside them. 
To overcome these limitations, 
in a previous analysis 
of the  quiet  chromosphere, we have explored the local emission 
at 17, 212, and 405 GHz frequencies 
and found that there are  
two 
main
regions of emission, the first one is close to the photosphere and the second
is located at $\sim 1000$ km over the photosphere
 \citep{2011ApJ...737....1D}. 

In 
the present
work we continue 
to  study the local radiative transfer 
processes in different 
layers of the atmosphere \citep[we present a brief description of the model in
  Section \ref{sec:model}, for details see][]{2011ApJ...737....1D}, 
for the empirical model SEL05 and two semi-empirical models: 
C7 and VALC (Section \ref{tprofi}),
but extend it to a larger range of frequencies, 
from 2 GHz to 10 THz (Section \ref{sec:synthetic}).
In particular, 
we study the radio emission associated with 
the  minimum of temperature and the lower chromosphere
(Section \ref{sec:cavity}).

It worth to mention that 
the study of the solar spectrum morphology, specially at
sub-millimeter and infrared
wavelengths, is also important 
for studies of dusty circumstellar
disks \citep{2001ARA&A..39..549Z} 
because
the emission from the chromosphere
of main-sequence solar-like stars at 
these
wavelengths could be comparable
with that emerging from circumstellar material.

\section{Radio Emission Model} \label{sec:model}
In order to compute the radio emission, we solve locally the radiative transfer
equation. 
The amount of energy $dI_\nu$ passing through volume element 
characterized by the 
length
$ds$ is equal to the absorption $I_\nu \kappa_\nu$
plus the emission $\epsilon_\nu$ inside the element, 
this 
is: 
\begin{equation}
\frac{dI_\nu}{ds} = -I_\nu \kappa_\nu + \epsilon_\nu
\end{equation}
where $\kappa_\nu$ is the opacity function and $\epsilon_\nu$ is the emission
function. 
The 
solution of the
radiative transfer equation can be represented in a local 
cell 
as follows:
\begin{equation}\label{emili}
I_{lcl} = \epsilon_{abs} + \epsilon_{emi},
\end{equation}
where $I_{lcl}$ is also know as the contribution function \citep{1976oasp.book.....G}, and
\begin{equation}\label{absi}
\epsilon_{abs} = I_0\exp(-\tau_{lcl})
\end{equation}
is the local absorption and
\begin{equation}\label{epsi}
\epsilon_{emi} = S_{lcl}(1-\exp(-\tau_{lcl}))
\end{equation} 
is the local emissivity.
In this case, $I_0$ is the incoming radiation,
$\tau_{lcl}$ is the local optical depth and
$S_{lcl}$ is the local source function.
We remind the reader 
that the specific intensity, $I_{\nu}$, can be converted to brightness
temperature, $T_b$, using the Rayleigh-Jeans approximation:
\begin{equation}
I_{\nu} = \frac{2k\nu^2T_b}{c^2},
\end{equation}
where $c$ is the speed of light and $k$ is the Boltzmann constant.
Then
we can re-write
equation (\ref{emili})
using
\begin{equation}
T_{blcl} =  \frac{I_{lcl}c^2}{2k\nu^2},
\end{equation}
where $T_{blcl}$ is the local brightness temperature.

In order to 
characterize
the 
local efficiency
emission and absorption of the atmosphere, 
we  define:
\begin{equation}\label{efiloca}
E_l = 1-\exp{(-\tau_{lcl})}
\end{equation}
as the local efficiency of emissivity and
\begin{equation}\label{absloca}
A_l = \exp{(-\tau_{lcl})}
\end{equation}
as the local efficiency of absorbance.
The values 
for
$E_l$ and $A_l$ can 
vary
between 0 and 1, where 0 means 
optically thin and 1 optically thick.

Using the local emissivity and absorbance, 
we compute iteratively the emission over a ray path by solving the 
radiative transfer equation \citep{2010ApJS..188..437D}.
Finally, 
the total emission efficiency, in the altitude range ($\xi$) from $h_1$ up 
to $h_2$, is: 
\begin{equation}\label{totemiefi}
\epsilon_T(h_1,h_2) = 1-\exp\left(-\sum_{\xi=h_1}^{h_2}\tau_{lcl}(\xi)\right) =  1- \exp(-\tau(h1,h2)) 
\end{equation}
where 
$\tau(h1,h2)$ is the optical depth between $h_1$ and $h_2$.

\section{The Solar Chromospheric Temperature Profiles}\label{tprofi}
The temperature structure in the Sun, and in general, in stars, decreases
monotonically with stellar (solar) radius. In the case of the Sun, more
elaborated models that include chromospheric heating
\citep[][VALC, SEL05, and C7 model respectively]{1981ApJS...45..635V,2005A&A...433..365S,2008ApJS..175..229A} indicate
that the temperature reaches a minimum  value at about 400 and 600 km above the solar
photosphere (Figure \ref{minimumTemperatures.eps}). Above this point, the
temperature gradually increase up to  coronal temperatures ($\approx$ 1 MK). 


A comparison among the three above mentioned models show only marginal
differences in the temperature profile, but quite significant discrepancies at
greater heights, including the value of the temperature minimum, For instance,
the VALC model predicts that the temperature at the minimum ($T_{min}$) is
4170 K,
and is located at 515 km above the photosphere. Similarly the models C7 and
SEL05 place the minima at 560 ($T_{min}= 4400 K$) and 500 km ($T_{min}= 4407
K$), respectively. 

By inspection of Figure \ref{minimumTemperatures.eps} one can see that the
results from the three models intersect at about 900 km. Above this point, the
models behave differently: The SEL05 profile increases 
linearly with height until 7700 K, while the VALC model increases from 5800 to
7200 K but not linearly, C7 grows from 5800 to 6650 K where remains almost
constant for the next 800 km. Finally, the VALC model presents a plateau in
temperature (around 24500 K) between 2115 km and 2267 km.
 

In the Figure \ref{relativeError.eps} we plot the 
differences in the radial temperature profiles between the three models,
taking as a reference the C7 model. We found three heights where the
differences in temperature are higher than 6\%: around 400 km (6\% only the VALC
model), 750 km (8.5\%), and 1100 km (6\%). In these cases the differences
with the C7 model are significant but always less than 10\%. 


\section{Synthetic Spectrum}   \label{sec:synthetic}
We use PakalMPI \citep{2011ApJ...737....1D}, an update version of the Pakal
code \citep{2010ApJS..188..437D} to compute a synthetic spectrum
of the solar emission from  millimeter to infrared wavelengths (from 2 GHz to
10 THz). 
PakalMPI solves the radiative transfer equation with  
NLTE conditions, in a three dimensional  
geometry using a multiprocessor environment. 
As model of emission, we use ``Celestun'' 
which
includes
three opacity
functions in the continuum: Bremsstrahlung \citep{1986rpa..book.....R}, Neutral Interaction 
\citep{1988A&A...193..189J, 1996ASSL..204.....Z} and Inverse Bremsstrahlung
\citep{1980ZhTFi..50R1847G}. 
Celestun model, also includes  ionization stages of
 twenty ion species considering $H$, $H^-$, and $n_e$ in NLTE \citep[see][for more details]{2011ApJ...737....1D}. 
The input parameters of 
PakalMPI 
are physical quantities 
as  temperature, Hydrogen density and metalicity. 
These parameters are taken   
from an atmospheric model, in this case the VALC, SEL05, and C7 models 
 mentioned in the previous section. 

In the Figure \ref{relativeTb.eps} we shows the differences between
the final brightness temperature (synthetic spectrum) for the three
models taking as reference the C7 model. We show that below 90 GHz the
differences between C7 and the VALC models exceed 20\%. The comparison
between C7 and SEL05 show differences of 30\% or less in all the frequency
range under study, being around 8, 200 and 1200 GHz the points in frequency that
presents very similar brightness temperatures.

Figure \ref{totalTau.eps} show the computed optical depth ($\tau$ in equation
\ref{totemiefi}) for each model at 10, 40, 200, and 800 GHz. 
For the C7 and the VALC models, between 480 and 1000 km over the photosphere
(marked with vertical lines), the optical depth remains
roughly constant. 
In other words, 
$\tau_{lcl}$ is very small and the local
emissivity is negligible (equation \ref{epsi}).

Using the SEL05 model, the local optical depth presents two steps in the same region.
This is due to the fact that this model 
has
an artificial peak enhancement of 
the Hydrogen density 
which 
causes an 
increment of the opacity at  $\sim 600$ km over the 
photosphere (Figure \ref{caiuswrong}). This artificial peak also   
modifies the total and
local efficiency. 
This region with Hydrogen over-density is a consequence of 
the fully ionized gas approximation used in the SEL05 model,  
which is not applicable at the 
low chromosphere
\citep{2011ApJ...737....1D}. 

The Figure \ref{opticaldepth.ps}
show the contour plot of the total emission efficiency 
(equation \ref{totemiefi})
as a function of frequency and height for the C7 model.
In particular we are interested in the Interface between Optically Thick and
Thin region ($IOTT$).
The interface can be defined as the region in
height where the emission efficiency is $0.9>\epsilon_T>0.1$.

This region
is presented in the plot as 
the gradient ribbon between red (optically thick) and blue (optically thin)
colors. We want to remark that the final brightness temperature comes from the
total optical depth and is related with the total emission efficiency. The
analysis of the local emission and absorption process help us to understand
the local emission and absorption process although remains hidden in the
observations.  

In the case of the C7 model
one can see
that the 
$10$ GHz 
emission comes from a layer situated at about $2100$ km above the photosphere. 
Whereas the $100$ GHz source 
is located at $1500$ km above the photosphere. 
The $IOTT$  in the  
2 - 500 GHz frequency range
is located in   
a large  range of heights, 
between 
1000 and 2100 km
above the photosphere. 
On the other hand, 
inside the 500 - 2000 GHz frequency range,
the
$IOTT$ remains
roughly constant
with altitude,
from  500 to 1000 km. 
This is 
 a direct consequence of the optical depth 
``Plateau'' showed
in Figure \ref{totalTau.eps}.

The results for the VALC model 
are very
similar 
to these obtained using 
the C7 model. We can see a slowly decreasing of the $IOTT$ between
10 and 300 GHz, then an abrupt change around 400 GHz is observed. 

At low frequencies, both 
C7 and VALC models 
predict a very low, and therefore unrealistic, altitude of the emitting sources.
This is due to the fact that these models 
do not take into account
the Corona and the Transition Region.

For the SEL05 model, the effect of the Hydrogen over-density in the emission efficiency 
is seen as a rapid shift of the $IOTT$ towards higher frequencies at 
$\sim 600$ km of height over the
photosphere, where the peak of Hydrogen is located (Figure \ref{caiuswrong}). 
In this case, 
the inclusion of ad-hoc Coronal physical conditions 
produces an increment in the  altitudes of the low frequency sources ($< 30$ GHz), 
giving  more realistic results for this range of frequencies  than the models
without Corona (C7 and VALC models).

In the Figure 5 from \cite{2011ApJ...737....1D}, we can observe
the brightness temperatures computed with
these models and their comparison with the observations 
\citep[collected by][]{2004A&A...419..747L}. 
This figure show that SEL05  provides the best match (although their
approximation in the low chromosphere disagree with the semi-empirical models
predictions).  
The results of C7 and VALC models
presents opposite behaviors. At lower frequencies the predicted
brightness temperature differs of the observations, C7 show lower and  VALC
higher brightness temperatures in the same range of frequencies. 
This
situation is 
partially inverted around 400 GHz where both models shows brightness
temperature above the observed but C7 predicts higher brightness temperatures
than 
VALC. If we observe the Figure \ref{relativeError.eps},
a direct relation between the differences in the radial temperature models and
the final brightness temperatures can not be easy established.

\section{The Chromospheric Millimeter-wave Cavity}  \label{sec:cavity}

\cite{2010ApJS..188..437D,2011ApJ...737....1D} reported 
that the emission profiles, in the  40 to
400 GHz range, have two regions of enhanced local 
emissivity. Remarkably, no attention 
has been
paid 
early
to this interesting behavior.
For instance, in figure 2 from \cite{2008ApJS..175..229A}, we note that at 0.5
mm there are also “two regions of emission”, unfortunately the authors did not elaborate much on this result. 
In this work we present 
a detailed
study 
of this region. 

We show in Figure \ref{opticalcavity.ps} the contour plot of the  
local emission efficiency 
(equation \ref{efiloca}) as a function of frequency and
height for the C7 model, where 
$E_l$
varies from optically thin (blue) to optically thick (red) medium. 
There is a peculiar zone of low emissivity centered at about
$\sim 750$ km above the photosphere.
This atmospheric  zone forms a cavity starting  at  $\sim 40$ GHz; its 
width grows with frequency; and ending  at $\sim 400$ GHz.

Between 40 and 400 GHz the cavity, surrounded by 
two zones of enhanced local emission is clearly seen. 
At a given frequency 
within
this range (for instance 100 GHz), and
moving outward
from the photosphere, we first 
find
a region where the atmosphere
emits efficiently (in red color between 0 and 600 km), then there is a region where the atmosphere
becomes thinner (in blue color between  600 and 900 km) and finally, 
a second region of emission appears (red
color between 900 and 1300 km), i. e. we have crossed three times 
the Local Interface between Optically Thick and Thin region ($IOTT_l$), in
this case defined as the region in height where the local efficiency of
emissivity is $0.9 > E_l > 0.1$.
The $IOTT_l$ divided the synthetic spectrum in three regions: 
between 2 and 40 GHz (where cross one time through the $IOTT_l$),
the second region between 40 and 400 GHz (cross three times)
and between 400 and 10000 GHz (cross one time).

The VALC and SEL05 models shows the same characteristic.
The 
SEL05 model show a more
complex structure, 
in particular it has a
characteristic
peak around 600 km over the photosphere,
which has been discussed previously.

We call this 
region 
the Chromospheric Solar Millimeter-wave Cavity (CSMC).  
This is 
a region of the solar
atmosphere 
where the 
opacity at 
millimeter and sub-millimeter
wavelengths is 
reduced,
surrounded by two zones of enhanced opacity.
As far as we know, 
the presence of 
the CSMC
have
never 
been 
studied or 
previously mentioned
in the literature.

\section*{Discussion and Conclusions}

The optical thickness of the solar atmosphere depends of the opacity functions involved in
the emission/absorption processes. 
At millimeter wavelengths the important
processes are the neutral interaction and the classical Bremsstrahlung \citep{2011ApJ...737....1D}.
These mechanisms depends of the temperature and the electronic and ion density. 

In the chromospheric layers, close to the
temperature minimum, the
rate of ionization is low
and therefore, the number of interactions between electron and ion decreases, 
resulting 
in locally thin atmospheric layers (at millimeter wavelengths). 
Around this region, 
the temperature 
raises
producing free electrons and ions and 
therefore, these atmospheric layers become optically thick. 
The interplay between the optically thick and thin layers produces the
Chromospheric Solar Millimeter-wave Cavity. 

We knows \citep{2011ApJ...737....1D} that the first 500 km over the
photosphere the Neutral Interaction is the most important mechanism in
the emission and absorption process then, the Bremsstrahlung comes the major
contributor in the optical depth. The second region of emission in the CSMC
begin around 700 km over the photosphere, we can concludes that the neutral
interactions do not influence the resulting CSMC.  

The morphology of the CSMC is a model-based phenomenon. In this work, we have
used  two well known semi-empirical models (VALC and C7) and one ad-hoc model
(SEL05) and all of them showed the CSMC structure but with small changes in its
morphology. 
We think that in average the used models reproduce well the physical
conditions in the low chromosphere due that the computations in another
regions of the spectrum show consistence
\citep{1981ApJS...45..635V,2008ApJS..175..229A}. However, the differences in the
computed synthetic spectra at millimeter, sub-millimeter and infrared regions
should be taken into account not only to test the auto consistence of the
models but as tool to improve the radial temperature profile. 

The presence of
the CSMC show that the relationship between the theoretical radial profile of
temperature and the observed spectrum is complex, against the general idea of
a simple  linear relationship as suggested by \cite{2008ApJS..175..229A}.  We
think that the CSMC must be a intrinsic characteristic of the solar
chromosphere and could help to understand better
the brightness temperature spectrum from  microwave to sub-mm 
range. 

A comparison between the observations \citep[collected
  by][]{2004A&A...419..747L} and the brightness temperature  
computed with C7 and VALC, predicts higher brightness
temperatures \citep[Figure 5 from][]{2011ApJ...737....1D} in almost all the
frequency range, except in the interval between 90 and 400 GHz.  
The SEL05 model fix better the same observations, but in this
work, we shows that the SEL05 approximations causes inconsistencies in the
Hydrogen density profile.

The $IOTT$ and the $IOTT_l$ define different frequency regions and heights of
emission in the synthetic spectrum. The $IOTT$ present three regions: 2 - 500
GHz, 500 - 2000 GHz, and  2000 - 10000 GHz while $IOTT_l$ shows: 2 - 40 GHz,
40 - 400 GHz and 400 - 10000 GHz. 
Although the final brightness temperature is related
directly with the $IOTT$ the true responsible behind the shape spectrum is the
$IOTT_l$, i.e. the CSMC. 

We can concludes that to fix the high brightness temperature computed with C7
and VALC (after 400 GHz), we need decrease the local optical depth around the
heights of the second peak of local emission of the CSMC. The opacity function
at this altitudes  mainly depends of the Bremsstrahlung process, i.e. lower
temperatures than the published for C7 and VALC between 700 and 1500 km over
the photosphere could decrease the local opacities values and then decrease
the brightness temperature.  
In particular, we expect that the sources of emission more sensitive to the
CSMC are at frequencies between 400 and 600 GHz, where finish the second peak
of emission of the CSMC. We expect that the height of the source of emission
may vary between 500 and 1000 km over the photosphere, causing a large
dispersion in the spectrum at this frequencies. 

On the low part of the spectrum, the inclusion of Transition Region and Corona
could fix the low brightness 
temperatures predicted by C7 at frequencies lowers than 40 GHz. The VALC plateau
of temperature between 2115 and 2267 km over the photosphere is the
responsible of the high brightness temperature computed with this model, in
particular in the centimetric range. Sub-millimeter solar observations are
needed in order to confirm the  presence of the CSMC particularly between 400
and 600 GHz. 

Finally, as the CSMC is a direct consequence of the minimum
temperature layers (mostly of the gradual temperature rise in the low chromosphere), it could give new insights about the
structure of the solar atmosphere, and also it can be extended to the case of solar-like
stars.

\acknowledgments
Part of this work was supported by  UNAM-PAPPIT IN117309-3 and CONACyT 24879 
grants.
Thanks to the National Center of Super-computing in Mexico for allow us to use
there computer facilities, Dr. Emanuele Bertone and Dr. Miguel Chavez for useful comments. Thanks
to Professor Pierre Kaufmann, director of CRAAM - Centro de Radioastronomia e
Astrofisica Mackenzie, where part of this research was conducted. JPR thanks
CNPq agency (Proc. 305655/2010-8).

\bibliographystyle{apj}

\appendix

\begin{figure}[h!]
\begin{center}
\includegraphics[width=1.0\textwidth]{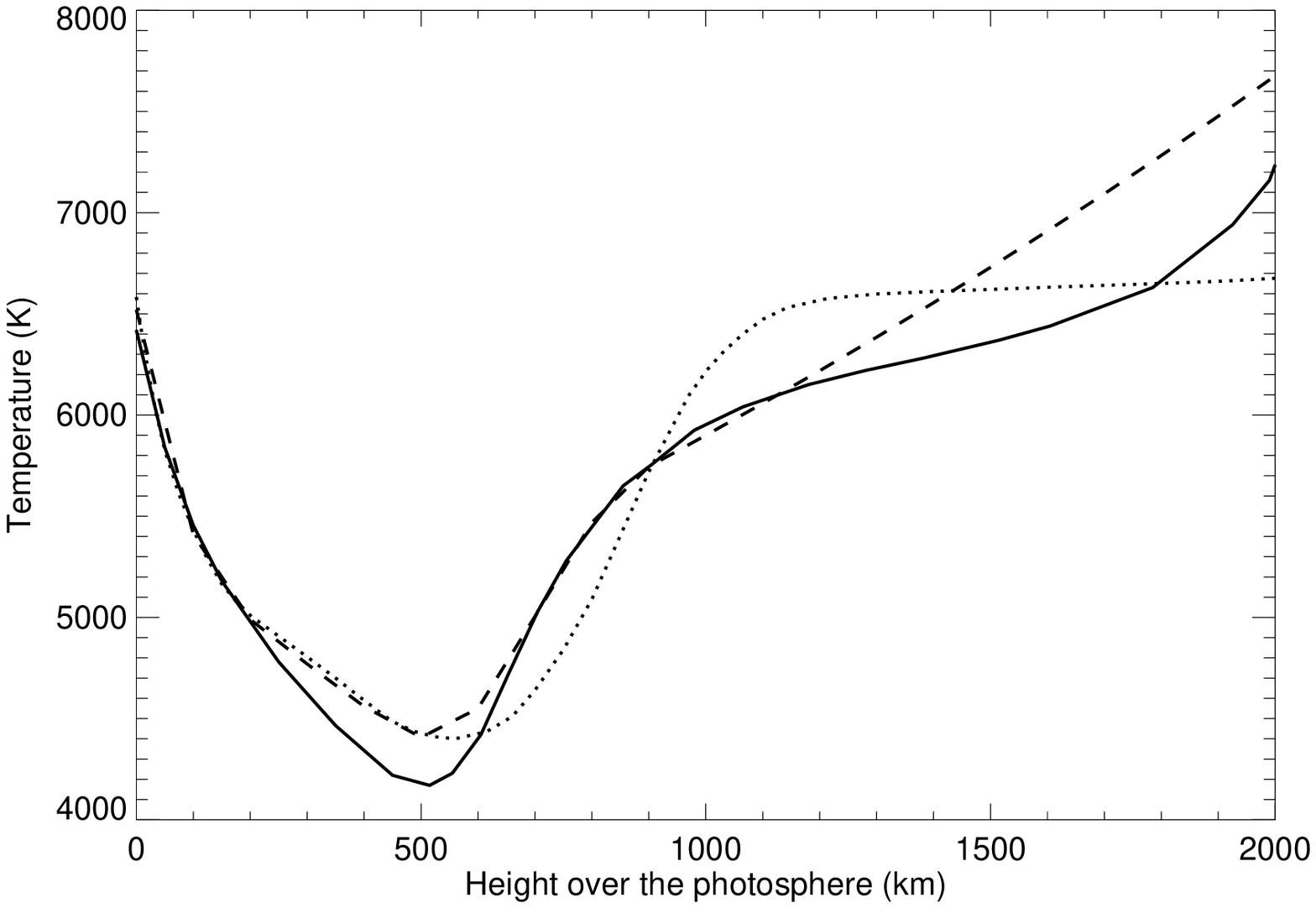}
\caption{
Radial temperature profiles of three Chromospheric models.  
The continuous 
line is the VALC model from\cite{1981ApJS...45..635V}, the dashed line corresponds to the  model of \cite{2005A&A...433..365S} and the dotted line is 
the C7 model
from \cite{2008ApJS..175..229A}.}\label{minimumTemperatures.eps} 
\end{center}
\end{figure}

\begin{figure}[h!]
\begin{center}
\includegraphics[width=1.0\textwidth]{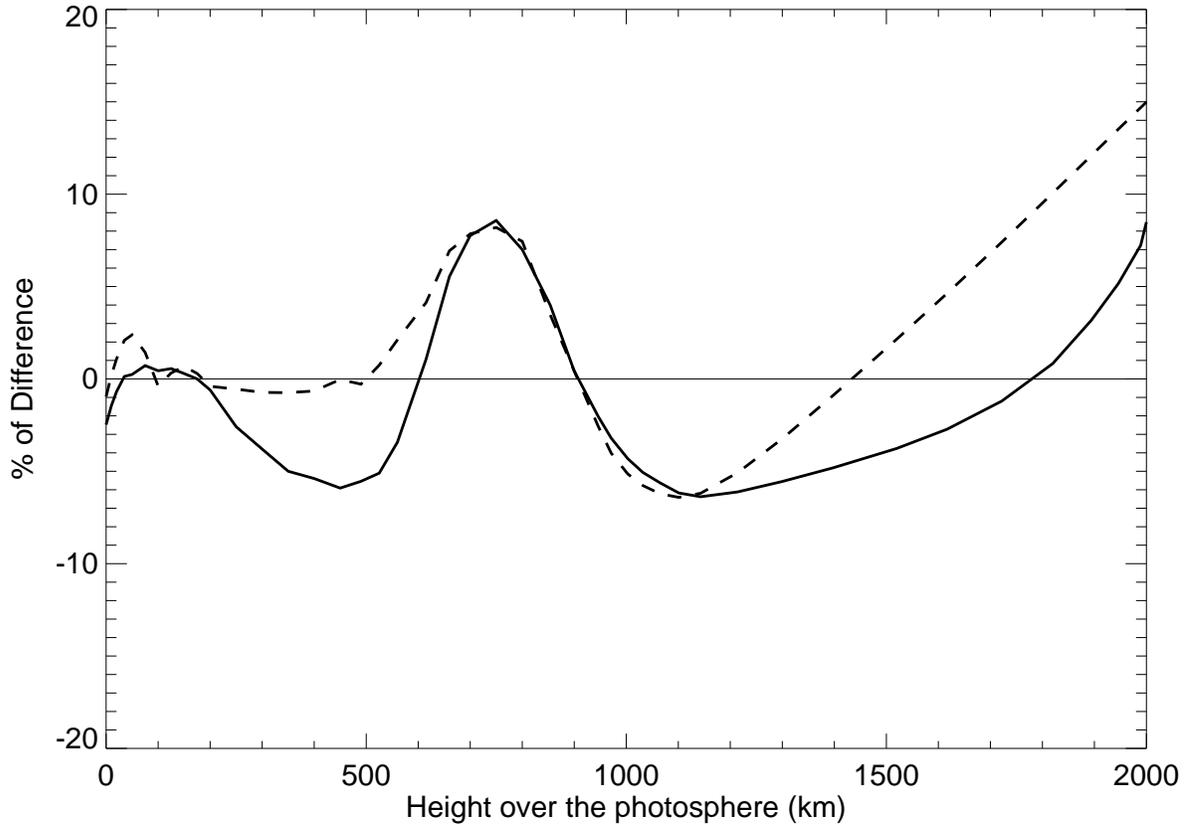}
\caption{Difference in the temperature profile between the C7 model (taking as reference) and the VALC (continuous line) and SEL05 (dashed line) models.
}\label{relativeError.eps} 
\end{center}
\end{figure}

\begin{figure}[h!]
\begin{center}
\includegraphics[width=1.0\textwidth]{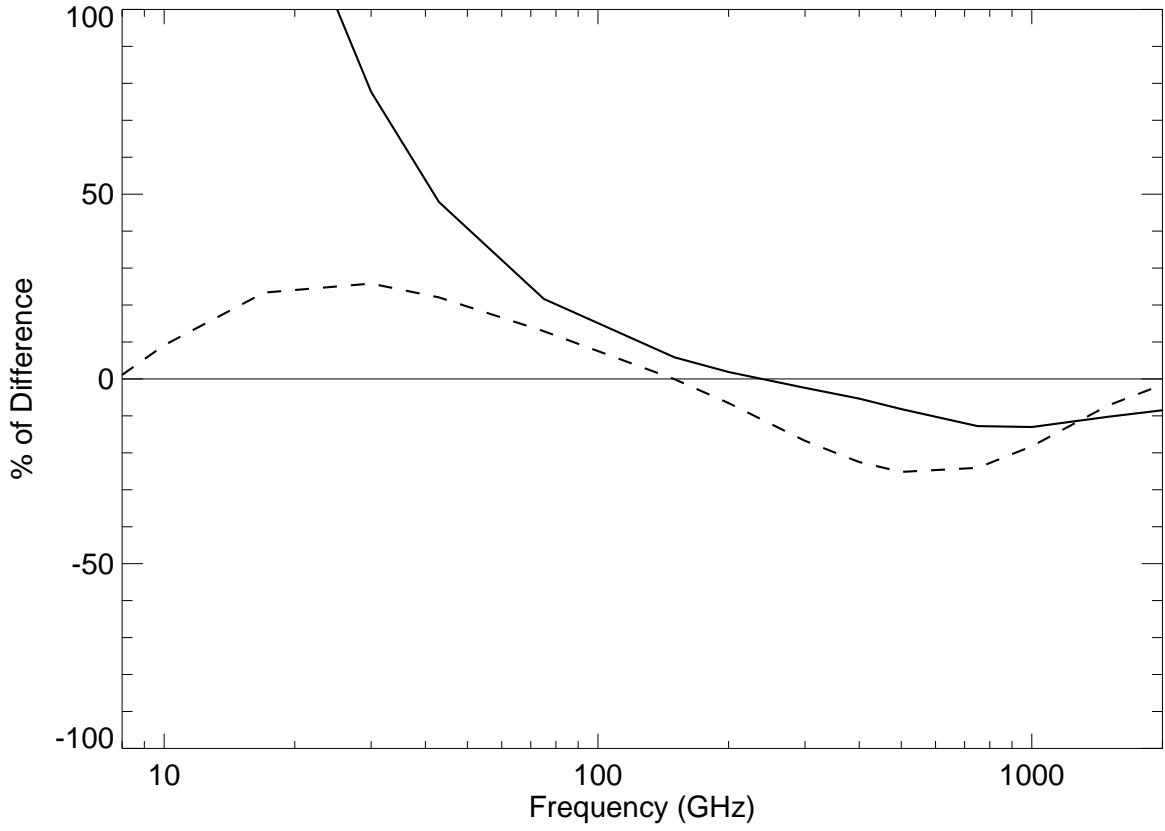}
\caption{Difference in the brightness temperature between the C7 model (taking as reference) and the VALC (continuous line) and SEL05 (dashed line) models.
}\label{relativeTb.eps} 
\end{center}
\end{figure}

\begin{figure}[h!]
\begin{center}
\includegraphics[width=0.8\textwidth]{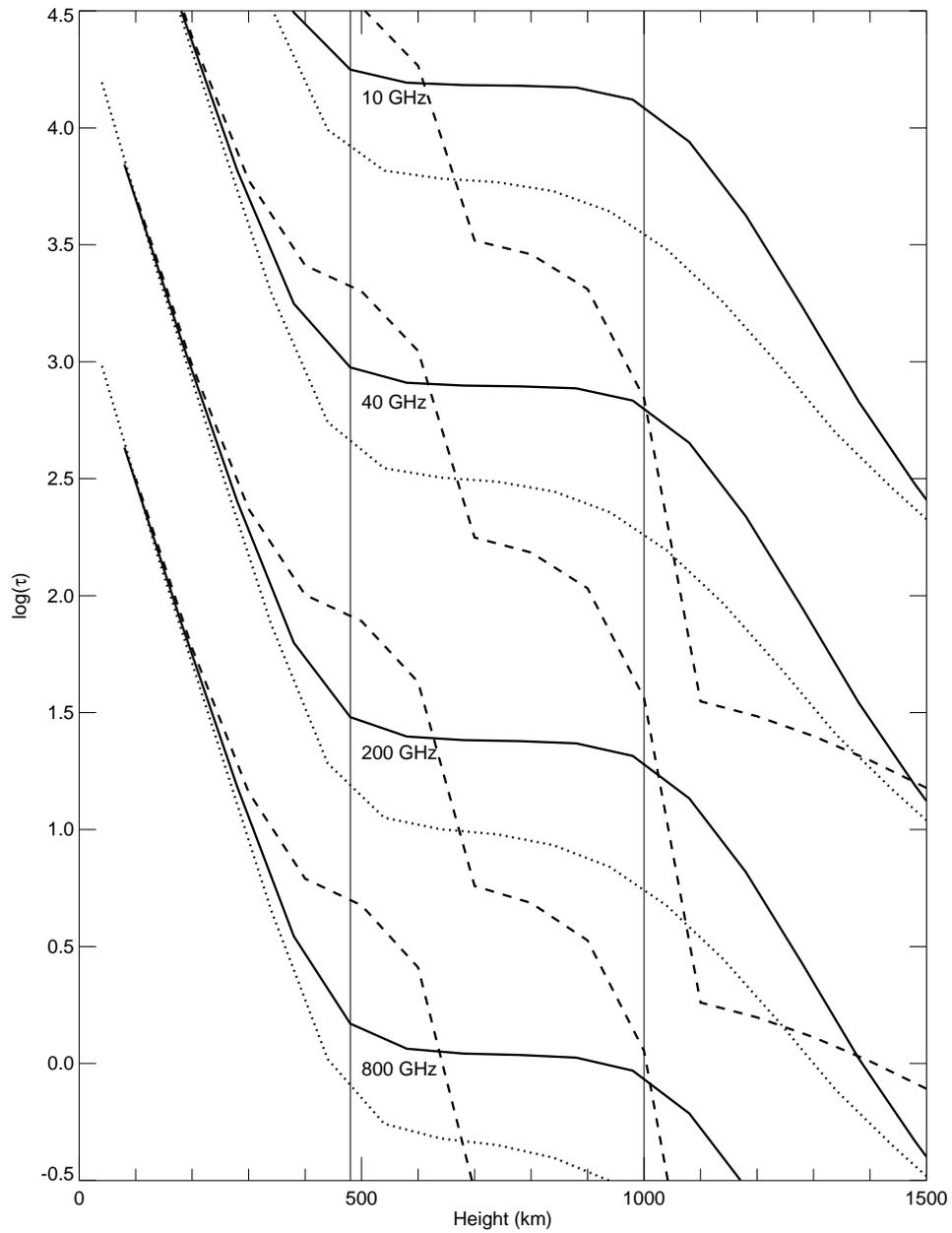}
\caption{Optical depth computed for 10, 40, 200 and 800 GHz using the C7
   (continuous line), VALC (dotted line), and SEL05  (dashed
  line) models. Note that the optical depth for the C7 and VALC models remains
  almost unalterable at heights between 480 and 1000 km over the photosphere, 
forming an
  optical depth Plateau.}\label{totalTau.eps} 
\end{center}
\end{figure}

\begin{figure}[h!]
\begin{center}
\includegraphics[width=1.0\textwidth]{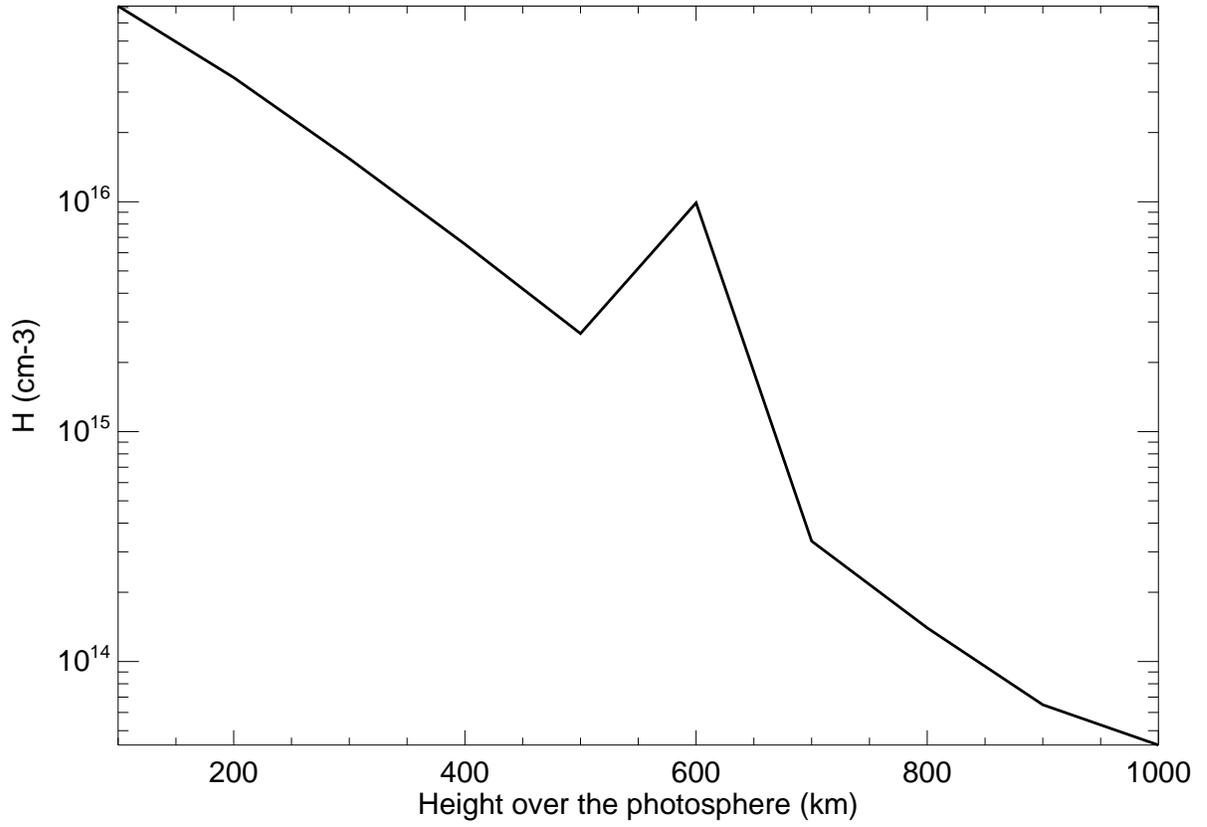}
\caption{Hydrogen density profile from SEL05 model. The peak around 550 km is 
a consequence 
of the full ionized gas hypothesis.}\label{caiuswrong} 
\end{center}
\end{figure}

\begin{figure}[h!]
\begin{center}
\includegraphics[height=0.8\textwidth,angle=270]{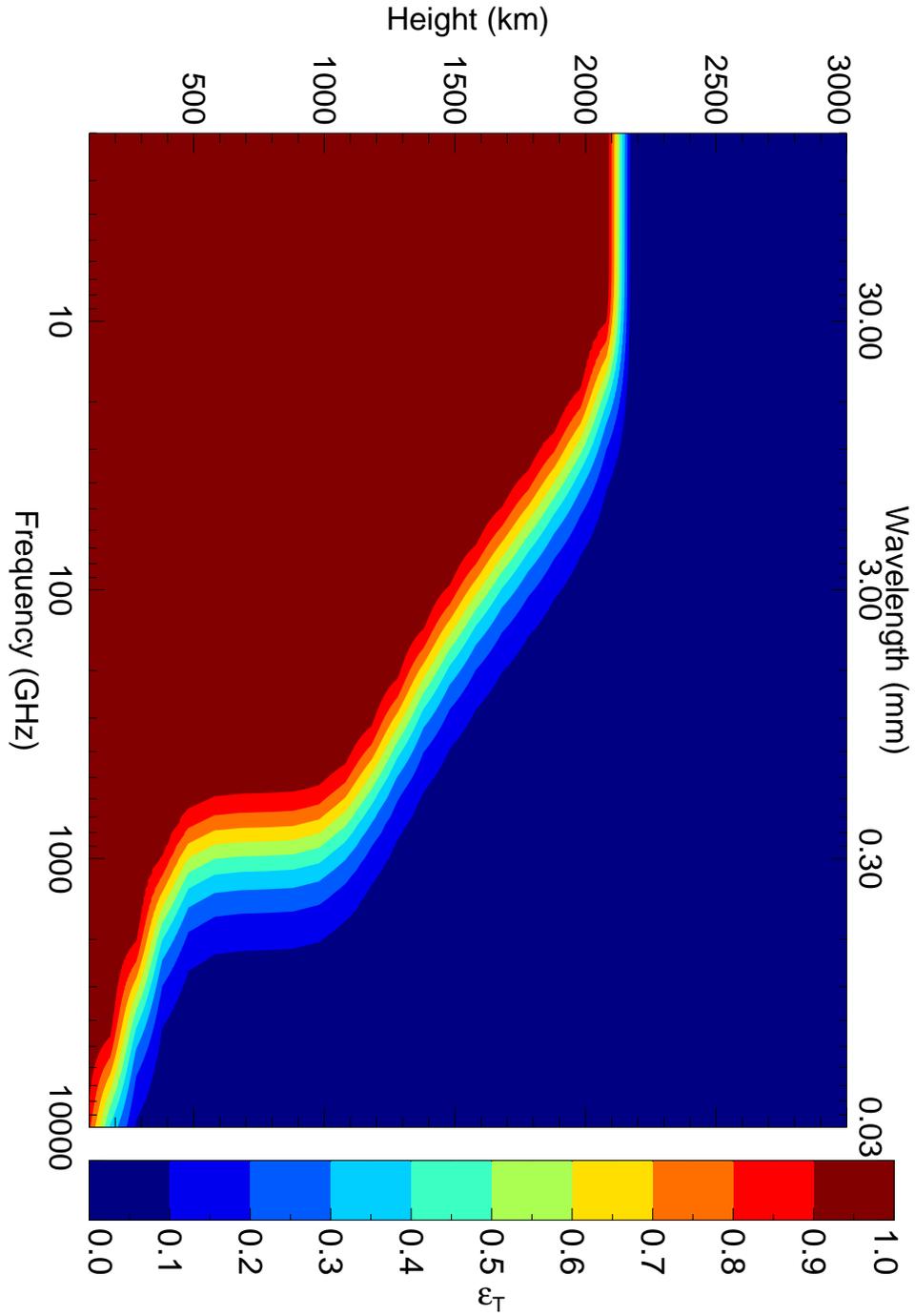}
\caption{Synthetic solar spectrum from 2 GHz to 10
  THz as a function of the atmospheric height over the photosphere. 
The contours correspond to  the efficiency of the
  total emission ($\epsilon_T$) computed using the C7 model
(for this and the following figures blue means optically thin and 
red optically thick). The region between 500 and 200 GHz is an effect of the
Optical Depth Plateau.}\label{opticaldepth.ps} 
\end{center}
\end{figure}

\begin{figure}[h!]
\begin{center}
\includegraphics[height=0.8\textwidth,angle=270]{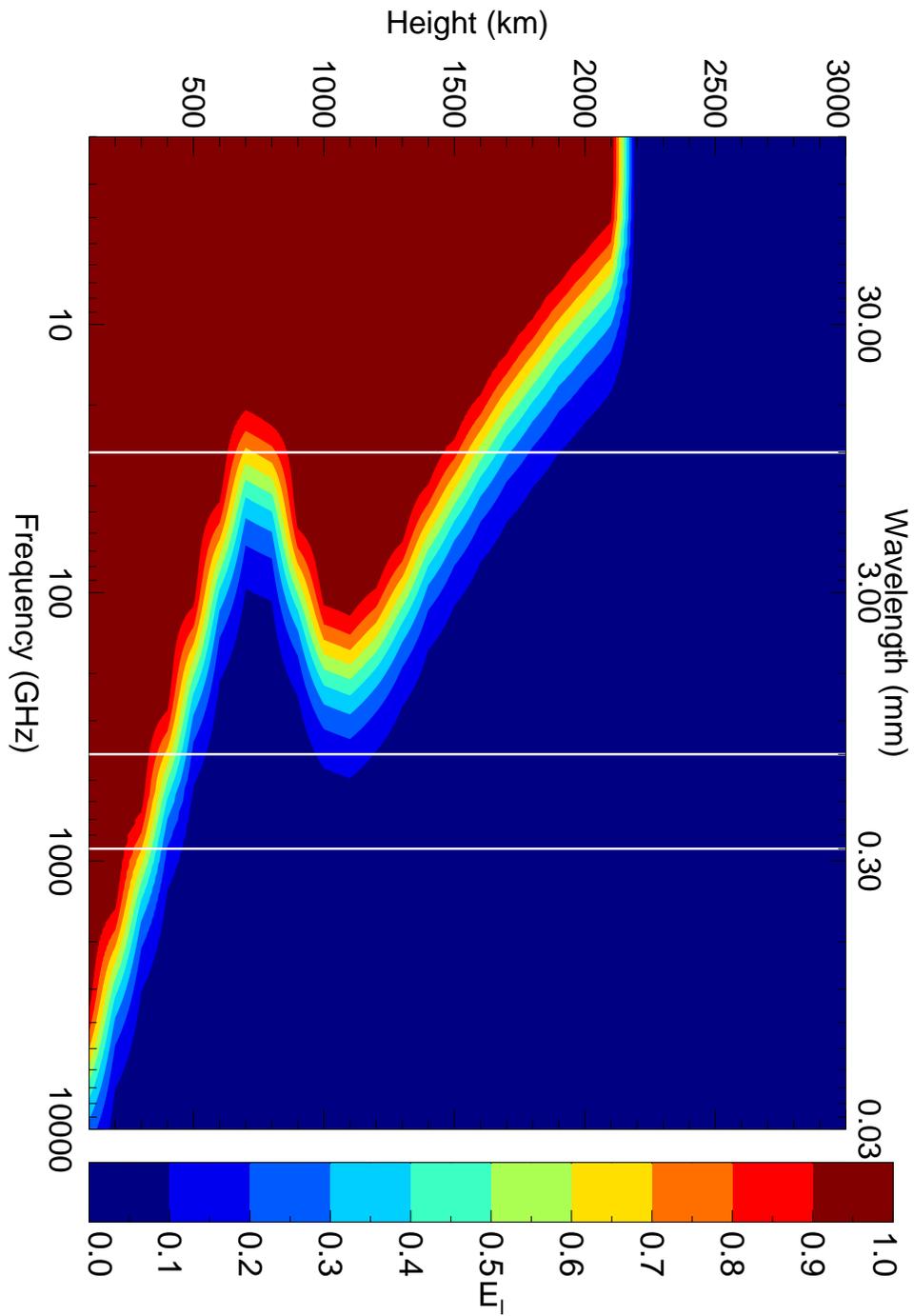}
\caption{Contour plot of the local  emissivity
  efficiency $E_l$ computed using 
the C7 model. 
the white vertical lines mark the Chromospheric Solar Millimeter-wave Cavity
(CSMC). 
}\label{opticalcavity.ps}
\end{center}
\end{figure}

\end{document}